\documentclass{article}
\usepackage{arxiv}
\usepackage[utf8]{inputenc} 
\usepackage[T1]{fontenc}    
\usepackage{url}            
\usepackage{booktabs}       
\usepackage{amsfonts}       
\usepackage{nicefrac}       
\usepackage{microtype}      
\usepackage{lipsum}		
\usepackage{graphicx}
\usepackage{natbib}
\usepackage{doi}
\usepackage{framed,multirow}
\usepackage{amssymb}
\usepackage{latexsym}
\usepackage{url}
\usepackage{xcolor}
\usepackage{hyperref}       

\usepackage{andrew_academic_paper}

\title{JOSA: Joint surface-based registration and atlas construction of brain geometry and function}

\author{Jian~Li\txtsup{1,2}, Greta~Tuckute\txtsup{3,4}, Evelina~Fedorenko\txtsup{3,4,5}, Brian~L.~Edlow\txtsup{1,2},\\ Adrian~V.~Dalca\txtsup{1,6,*}, Bruce~Fischl\txtsup{1,6,}\thanks{Co-senior authors with equal contribution}}
\affiliation{\txtsup{1} A. A. Martinos Center for Biomedical Imaging, MGH \& HMS\\
\txtsup{2} Center for Neurotechnology and Neurorecovery, MGH \& HMS\\
\txtsup{3} Department of Brain and Cognitive Sciences, MIT\\
\txtsup{4} McGovern Institute for Brain Research, MIT\\
\txtsup{5} Program in Speech Hearing Bioscience and Technology, Harvard University.\\
\txtsup{6} Computer Science and Artificial Intelligence Laboratory, MIT}

\date{}

\renewcommand{\headeright}{}
\renewcommand{\undertitle}{}
\renewcommand{\shorttitle}{J. Li \textit{et al.} JOSA}

\hypersetup{
pdftitle={JOSA: Joint surface-based registration and atlas construction of brain geometry and function},
pdfsubject={eess.IV, cs.CV, q-bio.NC},
pdfauthor={Jian~Li, Greta~Tuckute, Evelina~Fedorenko, Brian~L.~Edlow, Adrian~V.~Dalca, Bruce~Fischl},
pdfkeywords={Cortical Registration, Semi-supervised Learning},
}

\begin{document}
\maketitle

\begin{abstract}
  Surface-based cortical registration is an important topic in medical image analysis and facilitates many downstream applications. Current approaches for cortical registration are mainly driven by geometric features, such as sulcal depth and curvature, and often assume that registration of folding patterns leads to alignment of brain function. However, functional variability of anatomically corresponding areas across subjects has been widely reported, particularly in higher-order cognitive areas. In this work, we present JOSA, a novel cortical registration framework that jointly models the mismatch between geometry and function while simultaneously learning an unbiased population-specific atlas. Using a semi-supervised training strategy, JOSA achieves superior registration performance in both geometry and function to the state-of-the-art methods but without requiring functional data at inference. This learning framework can be extended to any auxiliary data to guide spherical registration that is available during training but is difficult or impossible to obtain during inference, such as parcellations, architectonic identity, transcriptomic information, and molecular profiles. By recognizing the mismatch between geometry and function, JOSA provides new insights into the future development of registration methods using joint analysis of the brain structure and function.\vspace*{2mm}
\end{abstract}

\keywords{Cortical Registration \and Semi-supervised Learning}

\section{Introduction}
\label{sec:intro}
Image registration is a key research topic in medical image analysis. Deformable image registration establishes spatial correspondence between a pair of images via a nonlinear spatial transformation. This transformation is often obtained through an optimization procedure that maximizes a similarity measure. Approaches for registration of the brain are either directly based on 3D images from scans of the brain (volume-based), or based on features resampled onto the brain surface (surface-based) \citep{Maintz_1998_MIA_SurveyMedical}, or a combination of these \citep{Postelnicu_2009_ITMI_CombinedVolumetric,Joshi_2009_IPiMI_FrameworkBrain}. Surface-based registration of the human cerebral cortex, often referred to as cortical registration, extracts a variety of representative information from the brain images and solves the registration problem based on a surface-matching framework.

Cortical registration methods align the complex folding patterns of the cerebral cortex by matching the geometric features of the brain \citep{Davatzikos_1996_ITMI_UsingDeformable,Fischl_1999_HBM_HighresolutionIntersubject}. Inter-subject cortical registration can improve not only the statistical power of group functional analyses \citep{vanAtteveldt_2004_N_IntegrationLetters,Frost_2012_N_MeasuringStructural}, but also the predictability of cytoarchitecture of the brain \citep{Yeo_2010_ITMI_LearningTaskoptimal,Fischl_2008_CC_CorticalFolding}. However, cortical registration is a challenging task because of the geometric complexity of the cortex and the large variability among individuals. Inter-subject surface alignment is commonly driven by geometric features that describe measures of cortical folding patterns, such as sulcal depth and mean curvature \citep{Fischl_1999_HBM_HighresolutionIntersubject,Yeo_2010_ITMI_SphericalDemons,Conroy_2013_N_IntersubjectAlignment}. It is often assumed that an accurate registration of folding patterns will also successfully align brain function \citep{Toga_2001_IaVC_RoleImage}. However, the functional variability of anatomically corresponding areas across subjects has been widely reported, particularly in regions implicated in higher-order cognition \citep{Fischl_2008_CC_CorticalFolding,Frost_2012_N_MeasuringStructural}. This variability implies that regions with different functional specializations may not be optimally aligned with a perfect anatomical registration. There may exist many equally good solutions for geometric registration with substantially different performance in registration of function. Another possible cause is the mismatch between geometry and function within each subject, i.e., different individuals may use a slightly different region to process the same function.

Traditional model-based deformable registration has been extensively studied \citep{Fischl_1999_HBM_HighresolutionIntersubject,Yeo_2010_ITMI_SphericalDemons,Sabuncu_2010_CC_FunctionbasedIntersubject,Guntupalli_2016_CC_ModelRepresentational,Robinson_2014_N_MSMNew,Vercauteren_2009_N_DiffeomorphicDemons,Nenning_2017_N_DiffeomorphicFunctional,Avants_2008_MIA_SymmetricDiffeomorphic,Beg_2005_IJCV_ComputingLarge,Christensen_1997_ITMI_VolumetricTransformation,VanEssen_1998_PNASU_FunctionalStructural,Joshi_2007_ITMI_SurfaceconstrainedVolumetric}. Typical strategies employ an iterative approach that seeks an optimal deformation field to warp a moving image to a fixed image. Methods usually involve optimization of a similarity measure between two feature maps, e.g., minimizing mean squared error (MSE) or maximizing normalized cross correlation, while regularizing the deformation field to have some desired property, such as smoothness and/or diffeomorphism. Widely used techniques for cortical surface registration map the surface onto the unit sphere and establish correspondence between feature maps in the spherical space \citep{Fischl_1999_N_CorticalSurfacebased}. Conventional approaches, such as FreeSurfer \citep{Fischl_1999_HBM_HighresolutionIntersubject}, register an individual subject to a probabilistic population atlas by minimizing the geometry MSE weighted by the inverse variance of the atlas convexity, in a maximum \textit{a posteriori} formulation. These anatomical registration methods have been adapted to functional registration by minimizing MSE on functional connectivity computed from functional magnetic resonance imaging (fMRI) data \citep{Sabuncu_2010_CC_FunctionbasedIntersubject}. Spatial correspondence can also be maximized by finding local orthogonal transforms that linearly combine features around each local neighborhood \citep{Guntupalli_2016_CC_ModelRepresentational, Guntupalli_2018_PCB_ComputationalModel}. Discrete optimization approaches iteratively align local features using spherical meshes from low-resolution to high-resolution \citep{Robinson_2014_N_MSMNew,Robinson_2018_N_MultimodalSurface}. To encourage invertibility of the deformation field and preserve anatomical topology, diffeomorphic registration uses an exponentiated Lie algebra, most often assuming a stationary velocity field (SVF) \citep{Ashburner_2007_N_FastDiffeomorphic, Vercauteren_2009_N_DiffeomorphicDemons}. These strategies were extended to the sphere by regularizing the deformation using spherical thin plate spline interpolation \citep{Yeo_2010_ITMI_SphericalDemons}. Several methods align functional regions, for example, using Laplacian eigen embeddings computed from fMRI data \citep{Nenning_2017_N_DiffeomorphicFunctional}. These methods are successful but solve an optimization problem for each image pair, resulting in a high computational cost.

Increasingly popular learning-based registration methods can be categorized into supervised \citep{Krebs_2017_MICCAI_RobustNonrigid,Sokooti_2017_MICCAI_NonrigidImage,Yang_2016_DLDLMA_FastPredictive,Cao_2017_MICCAI_DeformableImage} and unsupervised \citep{Dalca_2018_MICCAI_UnsupervisedLearning,Balakrishnan_2019_ITMI_VoxelMorphLearning,Cheng_2020_N_CorticalSurface,Niethammer_2019_PICCVPR_MetricLearning,Krebs_2019_ITMI_LearningProbabilistic,deVos_2019_MIA_DeepLearning} registration. Supervised registration evaluates the performance of a registration network by comparing the predicted deformation with a ``ground truth'' deformation. However, due to the lack of the ``ground truth'', deformation generated by iterative methods or simulation is often used as the target, which fundamentally limits the performance of supervised approaches \citep{Sokooti_2017_MICCAI_NonrigidImage,Yang_2016_DLDLMA_FastPredictive,Cao_2017_MICCAI_DeformableImage}. Further, the iterative registration methods are relatively slow which makes it impractical to generate large numbers of samples for supervised training, often resulting in overfitting. In contrast, unsupervised registration methods predict a deformation field that is used to warp the moving image to the fixed image. The core idea of these methods is to employ a classical loss in the image space, thus forming an end-to-end training pipeline \citep{deVos_2019_MIA_DeepLearning,Balakrishnan_2019_ITMI_VoxelMorphLearning,Hoffmann_2022_ITMI_SynthMorphLearning,Mok_2020_PICCVPRC_FastSymmetric}. Semi-supervised methods employ additional information, such as segmentation or parcellation maps, to guide registration without requiring them during inference \citep{Balakrishnan_2019_ITMI_VoxelMorphLearning,Blendowski_2021_MIA_WeaklysupervisedLearning}. Recent methods extend this strategy to the spherical domain by parameterizing the brain surfaces in a 2D grid that accounts for distortions \citep{Cheng_2020_N_CorticalSurface} or directly on the sphere using spherical kernels \citep{Zhao_2021_ITMI_S3RegSuperfast,Zhao_2021_ITMI_SphericalDeformable}. These learning-based registration methods substantially improved registration speed by orders of magnitude at inference while achieving a superior or comparable registration accuracy relative to the iterative methods. However, these methods do not directly model structure-function mismatches within a subject.

In this work, we present a diffeomorphic cortical registration framework that 1) explicitly uses functional data to drive the geometric registration to optimally align function, and 2) explicitly models mismatches between geometry and function. We build on recent unsupervised spherical registration strategies \citep{Balakrishnan_2019_ITMI_VoxelMorphLearning,Cheng_2020_N_CorticalSurface} and use a joint deformation field shared by geometry and function to capture the relatively large differences among subjects. We introduce deformation fields that describe relatively small variations between geometry and function within each subject. To avoid the biases of existing anatomical templates, we simultaneously learn a population-specific atlas during training \citep{Dalca_2019_ANIPS_LearningConditional}. We develop a semi-supervised training strategy that uses task fMRI data to improve functional registration but without requiring task fMRI data during inference. However, in contrast to the term ``unsupervised'' commonly used in the literature \citep{Balakrishnan_2019_ITMI_VoxelMorphLearning,Cheng_2020_N_CorticalSurface}, here we borrow the term ``semi-supervised'' to describe our strategy to highlight that auxiliary information, such as functional data, can be incorporated during training but are \textit{not} required during inference. This semi-supervised training framework can also be easily extended to any auxiliary data that could be helpful to guide spherical registration but is difficult or impossible to obtain or undesirable to use during inference, such as parcellations, architectonic identity, transcriptomic information, and molecular profiles. Therefore, the registration at inference is solely based on the geometry of the subject surface, which further avoids the potential circularity issue in any subsequent analyses of functional or other auxiliary data. We demonstrate experimentally that the proposed framework yields improved registration performance in both anatomical and functional domains.

We presented a preliminary version of this method at the Medical Imaging with Deep Learning 2023 conference \citep{Li_2023_PSCMIDL_JointCortical}. The current paper provides a more detailed description of the method, new and substantially expended experimental results with a comparison to an extensive list of state-of-the-art methods, as well as detailed ablation studies.

\section{Methods}
\label{sec:methods}

We describe \textbf{Jo}int \textbf{S}pherical registration and \textbf{A}tlas building (\textbf{JOSA}), a method for surface-based cortical registration with simultaneous atlas construction that explicitly models the cross-subject variation in the relationship between anatomy and functional properties in each subject.

\subsection{Generative model}
\label{sec:method:model}
\begin{figure}[bt]
  \centering
  \includegraphics[width=0.5\textwidth]{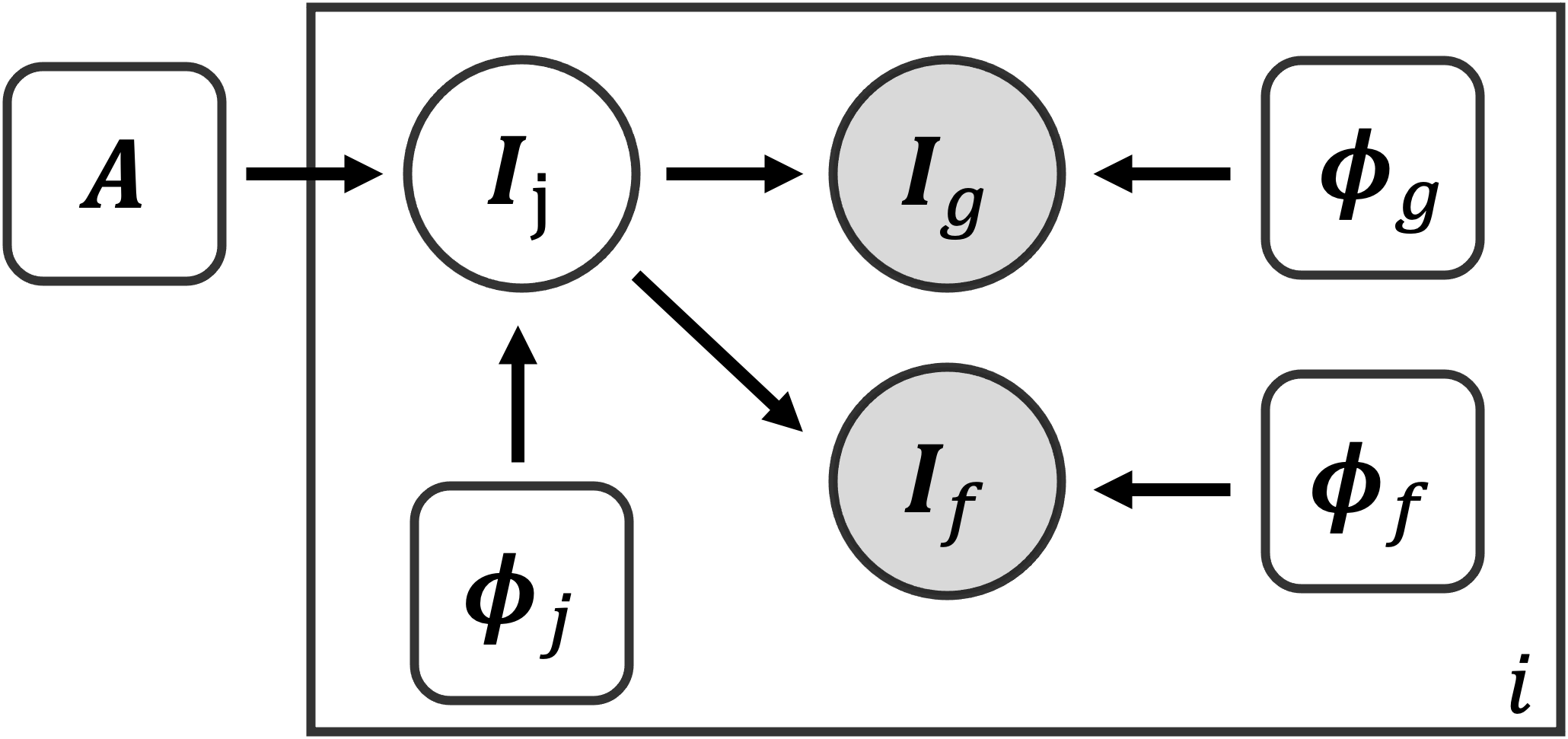}
  \caption{Graphical representation of the generative model. \textmd{Circles are random variables. Rounded squares indicate parameters. Shaded quantities are observations. The big plate represents replication. $\A$ represents the global atlas, $\I$ the input image, $\phiB$  deformation field. The subscript $j$, $g$, and $f$ stand for joint, geometry, and function, respectively.}}
  \label{fig:model}
\end{figure}

Fig.~\ref{fig:model} shows the graphical representation for the proposed generative model. Let $\A$ be an unknown population atlas with all geometric and functional cortical features of interest. We propose a generative model that describes the formation of the subject geometric $\I_{g}$ and functional  $\I_{f}$ features by first warping the atlas $\A$ by a subject deformation field $\phiB_{j}$. This model characterizes the differences between subjects and results in a joint multi-feature image $\I_{j}$. Geometric feature $\I_{g}$ is formed given an additional field $\phiB_{g}$ that deforms the geometric features in $\I_{j}$, and similarly for $\I_{f}$ and $\phiB_{f}$. The separation of $\phiB_{g}$ from $\phiB_{f}$ enables us to explicitly model structure-function variability across subjects. This is of critical importance from a neuroscientific perspective because some functional areas are better predicted by folding patterns than others \citep{Fischl_2008_CC_CorticalFolding}.

\subsubsection{Deformation priors}
\label{sec:method:prior}
 Let  $\phiB_{j}^{i}, \phiB_{g}^{i}, \phiB_{f}^{i}$ be the joint, geometric, and functional deformation fields for each subject $i$, respectively. All variables in the model are subject-specific, except for the global atlas $\A$, and we omit $i$ for our derivation. We impose the deformation priors
\begin{equation}
  \label{eq:prior}
  \begin{alignedat}{2}
    p(\phiB_{j}) & \sim &&  \text{ exp}\{-(\lambda_{j} \norm{\nabla \u_{j}}^{2} + \alpha_{j} \norm{\bar{\u}_{j}}^{2})\}\\
    p(\phiB_{g}) & \sim && \text{ exp}\{-\lambda_{g} \norm{\nabla \u_{g}}^{2}\}\\
    p(\phiB_{f}) & \sim && \text{ exp}\{-\lambda_{f} \norm{\nabla \u_{f}}^{2}\}
  \end{alignedat}
\end{equation}
where $\u_{j}$ is the spatial displacement for $\phiB_{j} = Id + \u_{j}$, $\nabla \u_{j}$ is its spatial gradient, and $\bar{\u}_{j}=1/N \sum_{i}\u_{j}^{i}$, $N$ is the number of subjects, and similarly for $\u_{g}$ and $\u_{f}$. The gradient term encourages smooth deformations, while the mean term encourages an unbiased atlas $\A$ by penalizing the average deformation over the entire dataset and hence favoring atlases that are ``close'' to every subject in the training. \citep{Dalca_2019_ANIPS_LearningConditional}.

\subsubsection{Data likelihood}
\label{sec:method:likelihood}

We treat the latent joint image $\I_{j}$ as a noisy warped atlas,
\begin{equation}
  \label{eq:lh-j}
  p(\I_{j} | \phiB_{j}; \A) = \gauss(\I_{j}; \phiB_{j}\circ \A, \sigma^{2}\mathbb{I})
\end{equation}
where $\gauss(\cdot; \muB, \SigmaB)$ is the multivariate Gaussian distribution with mean $\muB$ and covariance $\SigmaB$, $\circ$ represents spatial transformation, $\sigma$ represents additive noise, and $\mathbb{I}$ is the identity matrix. The geometric feature image $\I_{g}$ is then a noisy observation of a further-moved joint image~$\I_j$:
\begin{equation}
  \label{eq:lh-g}
  p(\I_{g} | \phiB_{g}, \I_{j}) = \gauss(\I_{g}; \phiB_{g}\circ \I_{j}, \sigma^{2}\mathbb{I}).
\end{equation}
Therefore, the complete \textbf{geometric image likelihood} is given by:
\begin{equation}
  \label{eq:lh-k}
    \begin{alignedat}{2}
      p(\I_{g} | \phiB_{g}, \phiB_{j} ; \A) &= \int_{\I_j} p(\I_{g} | \phiB_{g}, \I_{j}) p(\I_{j} | \phiB_{j}; \A)\\
      &= \int_{\I_j}  \gauss(\I_g; \phi_g \circ \I_j, \sigma \mathbb{I}) \gauss(\I_j; \phi_j \circ \A, \sigma \mathbb{I})\\
      &= \int_{\I_j}  \gauss(\I_j; \phi^{-1}_g \circ \I_g, \sigma \mathbb{I}) \gauss(\I_j; \phi_j \circ \A, \sigma \mathbb{I})\\
      &\stackrel{*}{=} \int_{\I_j}  \gauss(\I_j; \mathbf{\mu}_c, \mathbf{\Sigma}_c)  \gauss(\phiB^{-1}_{g}\circ\I_{g} ;   \phiB_{j} \circ \A; 2 \sigma^2 \mathbb{I})\\
      &=  \gauss(\phiB^{-1}_{g}\circ\I_{g} ;  \phiB_{j} \circ \A; 2 \sigma^2 \mathbb{I})\\
      &=   \gauss(\I_{g} ; \phiB_{g} \circ \phiB_{j} \circ \A; 2 \sigma^2 \mathbb{I})
    \end{alignedat}
\end{equation}
where in~$*$ we used an identity of the product of two Gaussian distributions, and~$\mu_{c}, \Sigma_{c}$ are constants. We use a similar model for the functional image~$\I_{f}$.
\label{sec:method:nn-semi-task}
\begin{figure*}[t]
  \centering
  \includegraphics[width=1\textwidth]{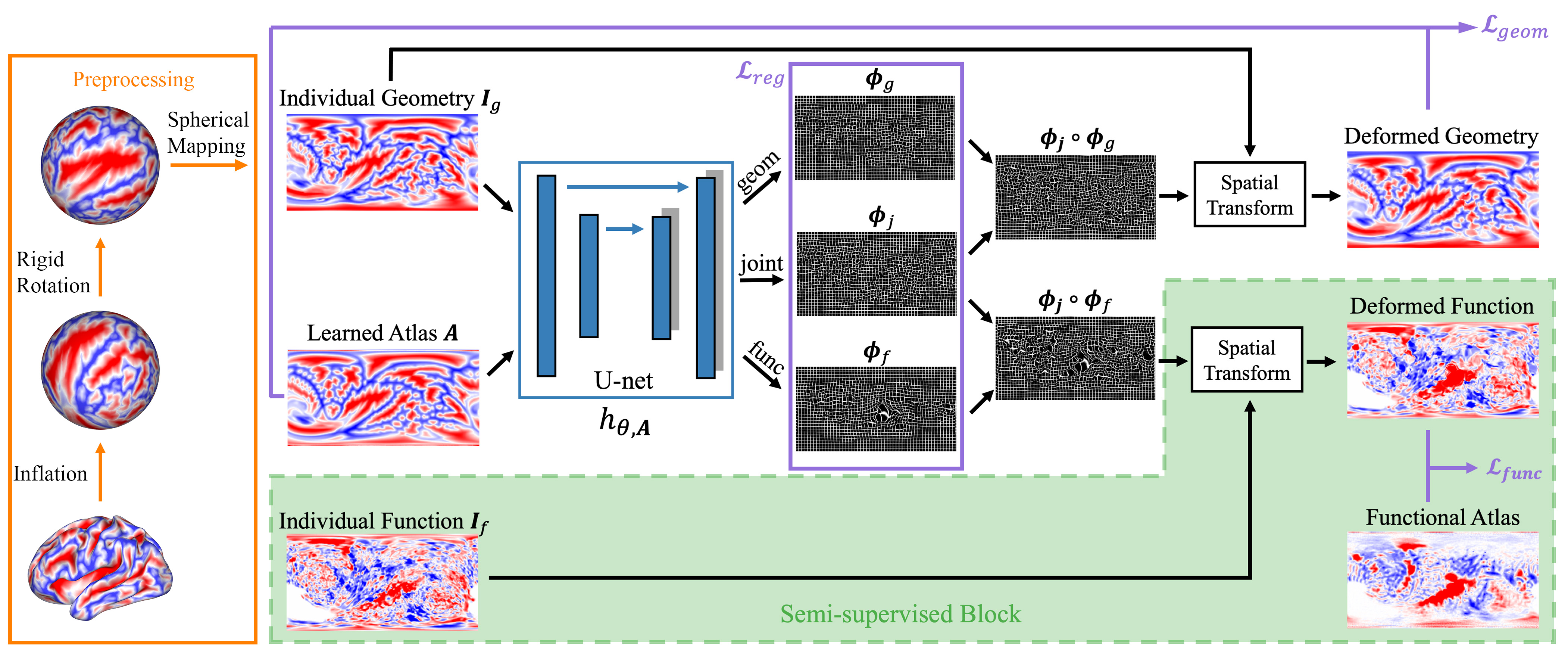}
  \caption{Network architecture and preprocessing pipeline. \textmd{The network takes the geometric features from the subject and outputs one joint and two separate deformation fields. The joint and geometric deformation are composted for registration of folding patterns, and likewise for function. Task fMRI data were used for evaluating the functional loss only in a semi-supervised manner, hence, functional task data are not required during inference. The losses $\mathcal{L}_{geom}$ and $\mathcal{L}_{func}$ represent the geometric part and the functional part of the data fidelity terms in Eq.~\eqref{eq:min-log}. They are evaluated in the atlas and subject space, which also helps avoid atlas drift \citep{Aganj_2017_N_MidspaceindependentDeformable} during atlas construction. $\mathcal{L}_{reg}$ represents the regularization and centrality terms which encourages smooth deformations and an unbiased estimation of the atlas.}}
  \label{fig:nn}
\end{figure*}

\subsubsection{Learning}
\label{sec:method:learning}
Let $\PhiB = \{\phiB_{j},\phiB_{g},\phiB_{f}\}$ and $\I = \{\I_{g},\I_{f}\}$. We estimate $\PhiB$ by minimizing the negative log likelihood
\begin{equation}
  \label{eq:min-log}
  \begin{alignedat}{4}
    \mathcal{L}(\PhiB | \I; \A) &= && -\log p(\PhiB | \I; \A) = -\log p(\I | \PhiB; \A) - \log p(\PhiB)\\
    &= && - \log   \prod_{k\in\{g,f\}}p(\I_{k}|\phiB_{j},\phiB_{k}, \A)  - \log  \prod_{k\in\{j,g,f\}}p(\phiB_{k}) \\
    &= &&\quad \frac{1}{2\sigma^{2}}\biggl(\norm{\I_{g}-\phiB_{g}\circ\phiB_{j}\circ  \A}^{2} + \norm{\I_{f}-\phiB_{f}\circ \phiB_{j}\circ  \A}^{2} \biggr)\\
    & && +  \lambda_{j}\norm{\nabla\u_{j}}^{2} + \lambda_{g}\norm{\nabla\u_{g}}^{2} + \lambda_{f}\norm{\nabla\u_{f}}^{2} +\alpha_{j}\norm{\bar{\u}_{j}}^{2}\\
    & && + \text{const.}
  \end{alignedat}
\end{equation}

The first two terms are data fidelity in the subject and atlas space respectively. The remaining four terms regularize the gradient and mean of the spatial displacement, encouraging smooth and unbiased deformation fields.

\subsection{Neural network and semi-supervised training strategy}
We use a neural network to approximate the function $h_{\theta, \A}(\I)=\PhiB$, where $\theta$ and $\A$ are network parameters. Fig.~\ref{fig:nn} shows the proposed network architecture. To work with surface-based data, the cortical surface of each subject is inflated into a sphere and then rigidly registered to an average space using FreeSurfer \citep{Fischl_2012_N_FreeSurfer}. Geometric and functional features are parameterized onto a 2D grid using a standard conversion from Cartesian coordinates to spherical coordinates, resulting in a 2D image for each input \citep{Cheng_2020_N_CorticalSurface}.

The network takes such a parameterized geometric image as input and outputs \textit{three} deformation fields in the parameterized coordinates. The joint deformation $\phiB_{j}$, which models the relatively large inter-subject variance, is shared among and composed with individual deformations $\phiB_{g}$ and $\phiB_{f}$.

We learn network parameters and use task fMRI data in a \textit{semi-supervised} manner. As shown in the green block in Fig.~\ref{fig:nn}, the task fMRI data and the corresponding functional atlas are not input into the neural network. Rather they are used only for evaluating the functional terms in the loss function~\eqref{eq:min-log}. This obviates the need for functional data during inference, as the deformation fields can be inferred using only geometric features. The proposed framework is flexible in the sense that auxiliary data, perhaps from different modalities, can be integrated into the framework for simultaneous multi-modality registration.

\subsection{Implementation}
\label{sec:method:implementation}
We implemented a Unet-like \citep{Ronneberger_2015_a_UNetConvolutional} network based on the core architecture in VoxelMorph (\url{https://voxelmorph.net}) \citep{Balakrishnan_2019_ITMI_VoxelMorphLearning,Dalca_2019_ANIPS_LearningConditional}. We used a 5-layer encoder with [128, 256, 384, 512, 640] filters and a symmetric decoder followed by 2 more convolutional layers with [64, 32] filters. Each layer involves convolution, max-pooling/down-sampling, and LeakyReLU activation. The spherical parameterization leads to denser sampling grids for regions at higher latitudes. To account for the difference in sampling density, we performed prior and distortion corrections described in SphereMorph\citep{Cheng_2020_N_CorticalSurface}. In short, weights proportional to $\sin(\theta)$, where $\theta$ is the elevation, were used to correct the distortion. SphereMorph also found that varying the locations of the poles in the projection had little impact on the resulting registration. The parameterized images were standardized identically but separately for structural and functional features, where the median was subtracted for each feature image followed by a division by the standard deviation.

During training, we randomly sampled the training data into mini-batches with a batch size of 8. For each batch, we augmented the data by adding Gaussian random deformations with a maximum $\sigma=8$ and a proper spherical topology and distortion correction at each spatial location. We further augmented the data by adding Gaussian noise with $\sigma=1$ for geometric features and $\sigma=6$ for functional features. We used the Adam optimizer \citep{Kingma_2014_AdamMethod} with an initial learning rate of $10^{-3}$. The learning rate was scheduled to decrease linearly to $10^{-4}$ within the first 500 epochs and then reduced by a factor of 0.9 if the validation loss does not decrease after every 100 epochs. The relative weights between functional loss and geometric loss were set to 0.7:0.3 empirically. We set the regularization hyperparameter $\lambda_{j}$ in Eq.~\eqref{eq:min-log} for the joint (large) deformation to be 0.1 and $\lambda_{g}$ and $\lambda_{f}$ for the individual (small) deformations to be 0.2. These hyperparameters were chosen where the validation loss is minimized.

The atlases, as part of the network parameters, were initialized using Gaussian random noise and automatically learned during training. We exploited a three-stage training strategy to further avoid atlas expansion or drift \citep{Aganj_2017_N_MidspaceindependentDeformable}: 1) we performed a training run using a single deformation with geometric features only, so that the network learned a geometric atlas; 2) we used and fixed the geometric atlas obtained in step one and trained a JOSA network to learn the functional atlas; and 3) we fixed both the geometric and the functional atlas from the previous two steps and performed a final training for the composite deformation fields.

We used TensorFlow \citep{Abadi_2016_TensorFlowLargeScale} with Keras front-end \citep{Chollet_2018_ASCL_KerasPython} and the Neurite package \citep{Dalca_2018_AnatomicalPriors}, and all experiments were conducted in a Dell Workstation with dual Intel Xeon Silver 6226R CPUs and an Nvidia RTX6000 GPU. The source code and the trained model of JOSA will be released to the public at \url{https://voxelmorph.net} as well as integrated as part of FreeSurfer \citep{Fischl_2012_N_FreeSurfer} package.

\section{Experiments}
\label{sec:experiments}

\subsection{Language task fMRI data}
\label{sec:data}
We used task fMRI data from a large-scale language mapping study \citep{Lipkin_2022_SD_ProbabilisticAtlas} ($N=800$), where the language network was functionally localized using a task that contrasts reading/listening of sentences versus a perceptually matched control (such as strings of nonwords or degraded speech), collected using a standard blocked design. Ten different versions of the language localizer task were used (Table 2 in \cite{Lipkin_2022_SD_ProbabilisticAtlas}), with the vast majority of subjects ($\sim77\%$) completing a version that contrasted reading of sentences with nonwords strings. In this localizer version, each stimulus was presented one word/nonword at a time at the rate of 450 ms per word/nonwords (12 words/nonwords per stimulus). Each stimulus was preceded by a 100 ms blank screen and followed by a 400 ms screen showing a picture of a finger pressing a button, and a blank screen for another 100 ms, for a total trial duration of 6s. Experimental blocks lasted 18 sec, and fixation blocks lasted 14 sec. Each run (consisting of 5 fixation blocks and 16 experimental blocks) lasted 358 sec. Subjects completed 2 runs. Subjects were instructed to read attentively and press a button whenever they saw the finger-pressing picture on the screen. Structural and functional data were collected on a 3T Siemens Trio scanner with a 12-channel (G1: $N=6$, G2: $N=12$) or a 32-channel (G3: $N=782$) head coil, at the Athinoula A. Martinos Imaging Center at the McGovern Institute for Brain Research at MIT. T1-weighted images were collected in 176 sagittal slices with 1 mm isotropic voxels. Functional data (BOLD) were acquired using an EPI sequence with the following parameters: 33 (G1) or 31 (G2, G3) 4 mm thick near-axial slices, 3.0 mm $\times$ 3.0 mm (G1) or 2.1 mm $\times$ 2.1 mm (G2, G3) in-plane resolution, FoV in the phase encoding (A $\ll$ P) direction 192 mm (G1) or 200 mm (G2, G3), matrix size 64 $\times$ 64 (G1) or 96 $\times$ 96 (G2, G3), TR = 2,000 ms, and TE = 30 ms.

We preprocessed the data using FreeSurfer v6.0.0 as described in \citep{Lipkin_2022_SD_ProbabilisticAtlas}. We sampled data onto the FreeSurfer average space (fsaverage), motion corrected, registered using the middle time point of each run, and spatially smoothed with a 4mm FWHM Gaussian filter. We reconstructed the subjects' surfaces from the T1 images (default \textit{recon-all} parameters). A ``sentence vs. nonword'' contrast t-map was originally generated for each subject using first-level GLM analysis based on the blocked design in the fsaverage space \citep{Lipkin_2022_SD_ProbabilisticAtlas}. We resampled the t-maps back onto the individual surfaces using FreeSurfer's `mri\_surf2surf` utility as functional features for registration in this work. We randomly split the data into a training set with 600 subjects, a validation set with 100 subjects, and a test set with the remaining 100 subjects.

\subsection{Baseline}
\label{sec:baseline}
We compare the registration performance of JOSA to FreeSurfer \citep{Fischl_1999_HBM_HighresolutionIntersubject}, Multimodel Surface Matching (MSM) \citep{Robinson_2014_N_MSMNew}, the improved version of MSM with higher-order constraint (MSM-HOC) \citep{Robinson_2018_N_MultimodalSurface}, and SphereMorph \citep{Cheng_2020_N_CorticalSurface} as surface registration baselines. For all baseline methods, we used three geometric features that quantify the folding patterns of the cortex to drive the registration, namely the mean curvature of the inflated surface (\verb|?h.inflated.H|), the sulcal depth map (\verb|?h.sulc|) and the mean curvature (\verb|?h.curv|) of the non-inflated white matter surface, all available as part of the FreeSurfer standard outputs. For FreeSurfer registration, we ran \verb|mris_register| (\url{https://surfer.nmr.mgh.harvard.edu/fswiki/mris_register}) for each test subject to register them to the fsaverage. For MSM and MSM-HOC, we ran the authors' pre-built \verb|msm| executable (\url{https://github.com/ecr05/MSM_HOC}) to register each subject to fsaverage with the configurations used in their 2014 and 2018 papers, respectively. For a fair comparison, both versions of MSM were executed in a hierarchical manner such that \verb|?h.inflated.H| was used for initial coarse alignment, then refined by \verb|?h.sulc| and \verb|?h.curv|, consecutively. For SphereMorph (\url{https://voxelmorph.net}), we used fsaverage as the registration target and trained the network to predict a single deformation field purely based on the three geometric features. We then used the predicted deformation to warp each subject's functional data to the fsaverage space at the inference time.

\subsection{Evaluation}
\label{sec:evaluation}
Qualitatively, we computed the group mean images of both the geometric and the functional data in the test set for each registration method. We then visualized them by superimposing the functional group mean map with the curvature group mean map.

\begin{figure}[tbh]
  \centering
  \includegraphics[width=0.7\linewidth]{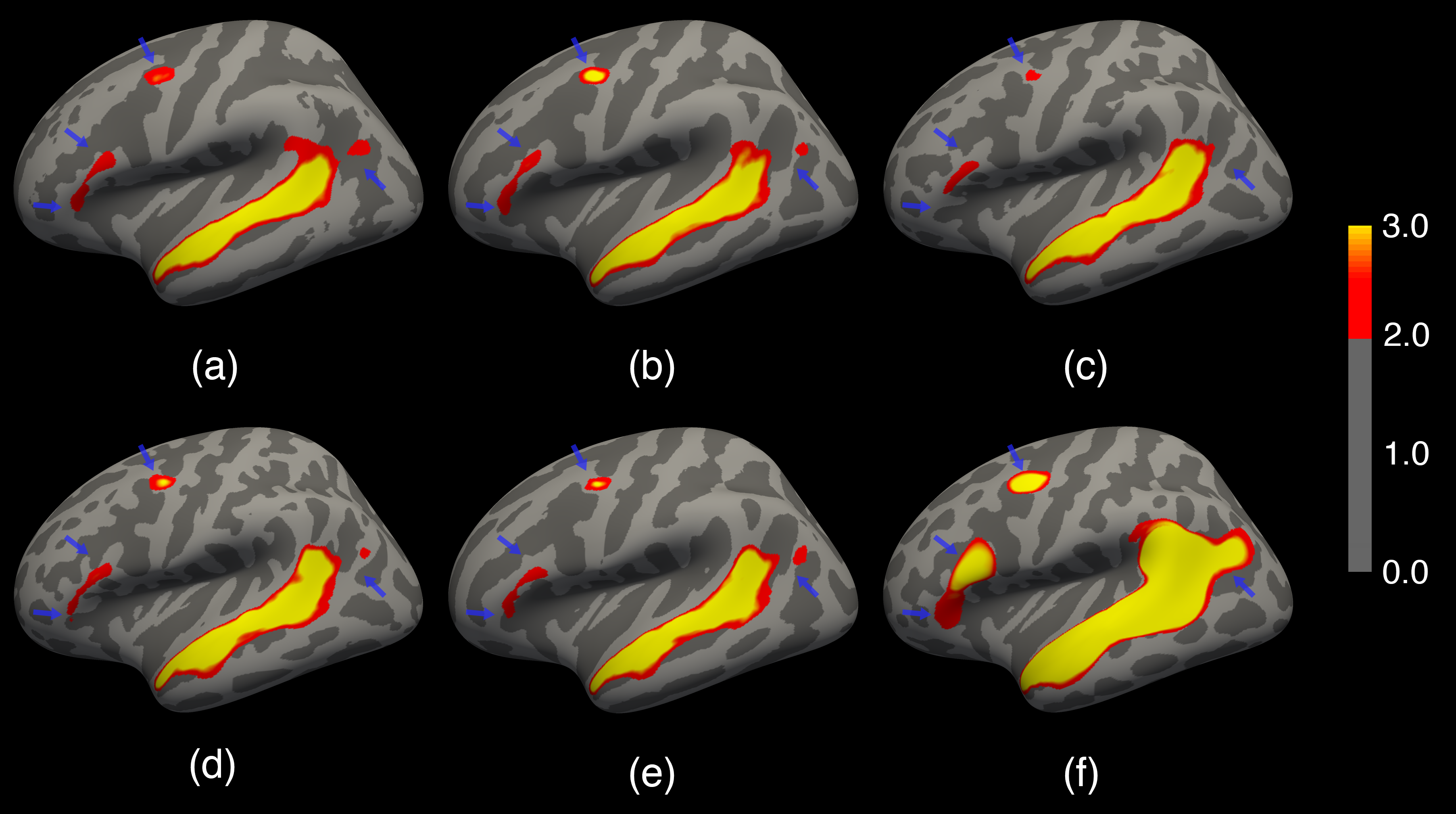}
  \caption{Group average curvature map of 100 test subjects overlaid with the group average functional language activation map. \textmd{(a) Rigid; (b) FreeSurfer; (c) MSM; (d) MSM-HOC; (e) SphereMorph; (f) JOSA. The language activation map is defined based on the contrast between well-formed sentences and a perceptually matched control stimulus. In the curvature maps, dark gray indicates sulci and light gray indicates gyri. The color bar for functional data (t-values) is shown on the right. Blue arrows highlight the key regions in functional alignment where large functional variability across subjects is often observed. JOSA achieved a substantially better alignment in both folding patterns and function with stronger functional response and sharper transition from task-active to non-active regions.}}
  \label{fig:qual}
\end{figure}

We quantified the geometric registration accuracy using the correlation between the registered individual data and the group mean \citep{Cheng_2020_N_CorticalSurface}. Specifically, we computed the Pearson correlation $c_{k}=corr(\I_{k}, \bar{\I})$ between the individual image $\I_{k}$ to the group mean image $\bar{\I}=1/N \sum_{k}\I_{k}$ for subject $k$, where $N$ is the number of subjects. We quantified the functional registration accuracy using the size of the overlapped regions between the suprathreshold individual functional maps and the suprathreshold group mean. We expect a better registration will have a larger surface area in the individual functional active regions as well as a larger overlap with the group mean for a reasonably conservative threshold. This also helps avoid a potential evaluation bias when a single-task contrast is used where only a fraction of the surface area is active during tasks. Specifically, we computed $s_{k}=|\I_{k}^{th}\bigcap \bar{\I}^{th}|$, where $\I_{k}^{th}$ is the suprathreshold individual image for subject $k$ at threshold $th$ and similarly for the group image $\bar{\I}^{th}$, and $|\cdot |$ is the measure of cardinality. We used an empirical threshold of $3$ in this evaluation based on the t-values in the language active regions. The suprathreshold regions for the group mean are shown in Fig.~\ref{fig:qual}.

To evaluate the consistency in registration improvement across subjects, we assessed the pair-wise geometric correlation difference $c_{Method}^{k} - c_{Rigid}^{k}$ between each of FreeSurfer, MSM, MSM-HOC, SphereMorph, and JOSA and the rigid alignment (i.e., the correlation difference for the \textit{same} subject after registration in comparison to the rigidly aligned version). We similarly assessed the pair-wise functional difference $s_{Method}^{k} - s_{Rigid}^{k}$.

In functional evaluations, to further investigate the impact of the selected threshold $th$, we computed the sum of the superthreshold t-values in the group mean $\sum_{j} \bar{\I}_{j}^{th}$ as a function of the threshold $th\in \{3, ..., 5\}$. We also looked into the two individual factors that have impact on the sum of the suprathreshold t-values, namely the number of vertices that have t-values above the threshold $M=|\bar{\I}^{th}|$ as well as the average of the suprathreshold t-values within those regions $1/M \sum_{j} \bar{\I}_{j}^{th}$. The former reflects the size of the functional active regions and the latter reflects the strength of the functional response.

We also compared the computational efficiency of the registration methods by measuring the run time of the registration procedure for each of the test subjects. Iterative methods (FreeSurfer, MSM, and MSM-HOC) were run on a single CPU thread, while the learning-based methods (SphereMorph and JOSA) were evaluated using both a CPU and a single GPU.

\subsection{Stability of atlas construction}
\label{sec:atlas-stability}
Atlas initialization can substantially bias the result \citep{Fischl_1999_HBM_HighresolutionIntersubject,Dickie_2017_FN_WholeBrain, Cheng_2020_MUPPPIA_UnbiasedAtlas}. To investigate the stability and the potential bias of atlas construction using JOSA, we repeated the geometric atlas learning procedure $6$ times, each time with randomly selected batches of subjects and their orders through the training process. We qualitatively visualized the difference between each run to the average of all runs. To quantitatively measure the stability of the learned atlases, for each pair of the learned atlases with the $6$ trials (hence $15$ pairs in total), we computed Pearson correlation between the pairs of atlases.

\subsection{Ablation studies}
\label{sec:ablation}
We conducted the following two ablation studies: 1) To illustrate benefit of learning the population-specific and unbiased atlas, we trained two JOSA networks with the only difference being in the registration target, where we used rigid average and FreeSurfer average for comparison to our learned atlas; 2) To assess the effect of modeling the difference between geometry and function, we trained an additional JOSA network with a single deformation field that was used to align both geometry and function for comparison to our two separate deformations.

\section{Results}
\label{sec:results}
\begin{figure}[b!]
  \centering
  \includegraphics[width=0.8\linewidth]{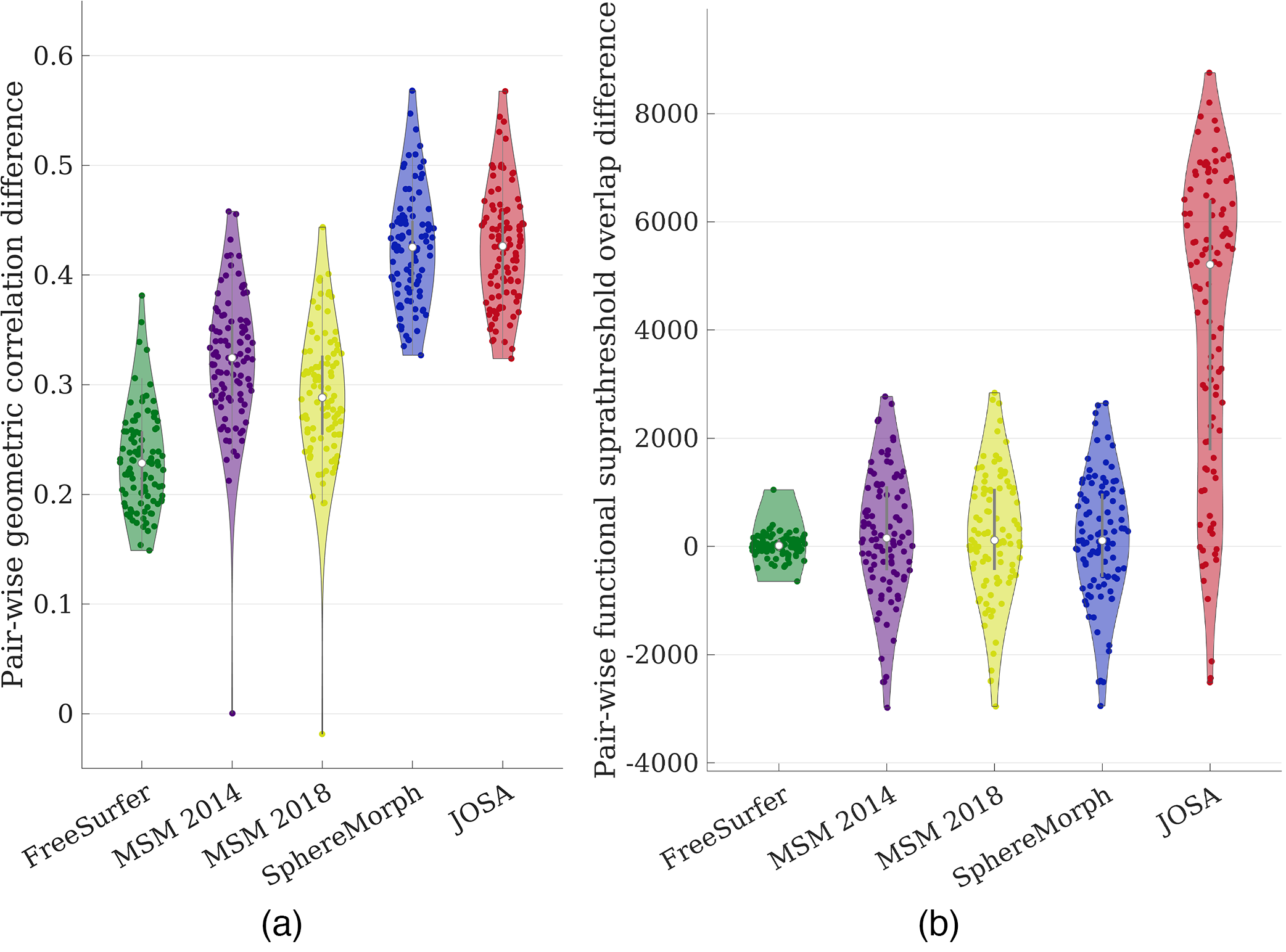}
  \caption{Violin plot of pairwise difference between each registration method to rigid alignment. \textmd{(a) Geometric difference measured by correlation; (b) Functional difference measured by the suprathreshold (t-values $>=3$) functional overlaps. JOSA yielded substantially higher correlation/larger overlap against rigid alignment compared to baseline methods and the improvement is consistent across subjects, as shown by the pairwise difference measures.}}
  \label{fig:violin}
\end{figure}

\subsection{Qualitative comparison}
\label{sec:res:qual}
Fig.~\ref{fig:qual} shows the group mean curvature maps (one of the geometric features) of the 100 test subjects overlaid with the group mean functional language activation maps for each of the registration methods. A better registration leads to a stronger average response (larger yellow areas) to language stimuli and a sharper transition from task-positive to task-negative regions (narrower margin from yellow to transparent regions). FreeSurfer, MSM, MSM-HOC, and SphereMorph significantly improved the registration of the folding patterns compared to the rigid alignment, albeit with some noticeable but minor differences in the average curvature patterns. However, all these methods only yield marginal improvement of the functional alignment. In contrast, JOSA achieved a substantially better alignment in both folding patterns and function. In particular, the predominant language region in the superior temporal gyrus shows a substantially stronger response and clearer functional boundaries. We also find additional/stronger language-responsive regions in the frontal eye field, the area in the inferior frontal gyrus (BA 44/45 near the classic Broca's area), as well as the posterior section of the inferior parietal lobule as indicated by the blue arrows, all consistent with the probabilistic language atlas in earlier work \citep{Lipkin_2022_SD_ProbabilisticAtlas}.

\subsection{Quantitative evaluation}
\label{sec:res:quan}
Quantitatively, Fig.~\ref{fig:violin}~(a) shows the pair-wise geometric correlation difference between each of the registration methods to the rigid alignment, and Fig.~\ref{fig:violin}~(b) shows the pair-wise functional suprathreshold overlap difference. Values higher than zero indicate consistent registration improvement over rigid alignment. Fig.~\ref{fig:orig_violin} in the supplementary material shows the original correlation/overlap values without taking pair-wise differences. Iterative methods such as FreeSurfer and MSM substantially improved geometric correlations over rigid alignment (by $\sim 0.3$). Learning-based methods achieved a further significant improvement (by $\sim 0.43$) and were comparable between SphereMorph and JOSA. In contrast, only JOSA substantially improved functional registration, while all other methods did not improve or only marginally improved functional registration compared to rigid alignment.

\begin{figure}[tb]
  \centering
  \includegraphics[width=0.55\linewidth]{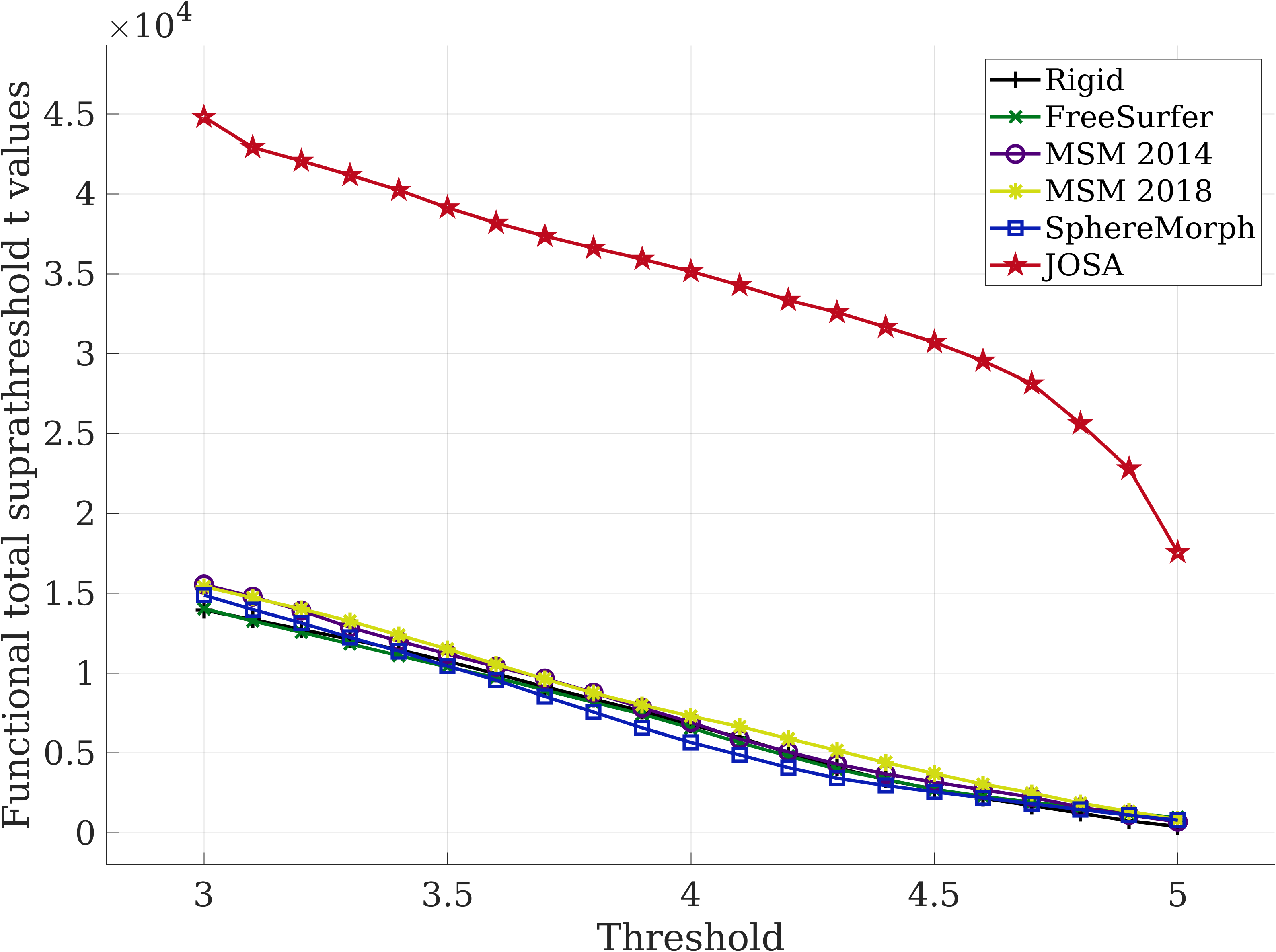}
  \caption{Functional total suprathreshold t-values of the group means as a function of the threshold. \textmd{The total suprathreshold t-values in JOSA is substantially higher than that using baseline methods regardless of the selected threshold.}}
  \label{fig:func_total}
\end{figure}

\renewcommand{\arraystretch}{1.3}
\begin{table}[b!]
  \caption{Comparison of the run time for individual subject registration. Time is measured in minute unless noted specifically\label{tab:run-time}}\vspace*{3mm}
  \centering
  \small
  \begin{tabular*}{0.7\linewidth}{@{\hspace{3mm}}L{3cm}@{}L{3cm}@{}L{3cm}@{}L{3cm}@{}}
    \hline
    Method & CPU run time & GPU run time & GPU run time\newline inference only\\ \hline 
    FreeSurfer  & $8.82\pm 2.47$ & \\
    MSM & $21.05\pm 1.87$ & \\
    MSM-HOC & $57.09\pm 8.86$ & \\
    SphereMorph & $0.097\pm 0.008$ & $0.095\pm 0.008$ & $0.003\pm 0.004$\\
           & $5.81\pm 0.49$ sec & $5.68\pm 0.51$ sec & $0.18\pm 0.26$ sec\\
    JOSA & $0.097\pm 0.009$ & $0.094\pm 0.008$ & $0.003\pm 0.003$\\
           & $5.83\pm 0.53$ sec & $5.67\pm 0.49$ sec & $0.17\pm 0.09$ sec\\ \hline
  \end{tabular*}
\end{table}
\renewcommand{\arraystretch}{1}

Fig.~\ref{fig:func_total} shows the total suprathreshold t-values of the group mean functional image for each registration method as a function of the threshold. All geometry-based registration methods yielded similar functional total t-values to that using only rigid alignment. In contrast, JOSA produces substantially higher functional total t-values. This significant difference between JOSA and baseline methods was consistent regardless of the choice of the threshold. The higher total t-values may come from either a larger region where the functional values exceeded the threshold, a stronger response, or both. To further assess the effect, we plotted the curves for (a) the number of vertices within regions that have functional values above the threshold and (b) the averaged functional values within those regions in the Supplementary Fig.~\ref{fig:vert_strength}. JOSA has both larger functional regions as well as a stronger functional response within those regions, and the former contributes more than the latter.

\subsection{Run time}
\label{sec:res:runtime}
Table~\ref{tab:run-time} summarizes the run time for each registration method. For a single subject, all methods started with the spherical cortical surface, and the output is the warped spherical surface. The default FreeSurfer takes approximately $9$ min to complete the spherical registration procedure. MSM and MSM-HOC operated directly on the sphere without parameterization but took significantly longer to execute. In contrast, the learning-based methods take, on average, only $0.097$ min ($\approx 5.8$ sec!), providing a speed up by \textbf{two orders of magnitude} compared to the iterative methods. The registration speed can be further accelerated by \aprx $0.15$ sec per subject with a GPU implementation. This $0.15$ sec CPU-to-GPU improvement is significant because, in both SphereMorph and JOSA, the parameterization and resampling back to sphere take most of the time during registration, and the actual inference time through the convolutional neural network is \aprx $0.17$ sec using GPU (last column). That means if the parameterization and resampling can be parallelized and precomputed on a computer cluster, registering $1000$ subjects can be done in under $3$ minutes.

\begin{figure}[b!]
  \centering
  \includegraphics[width=0.5\linewidth]{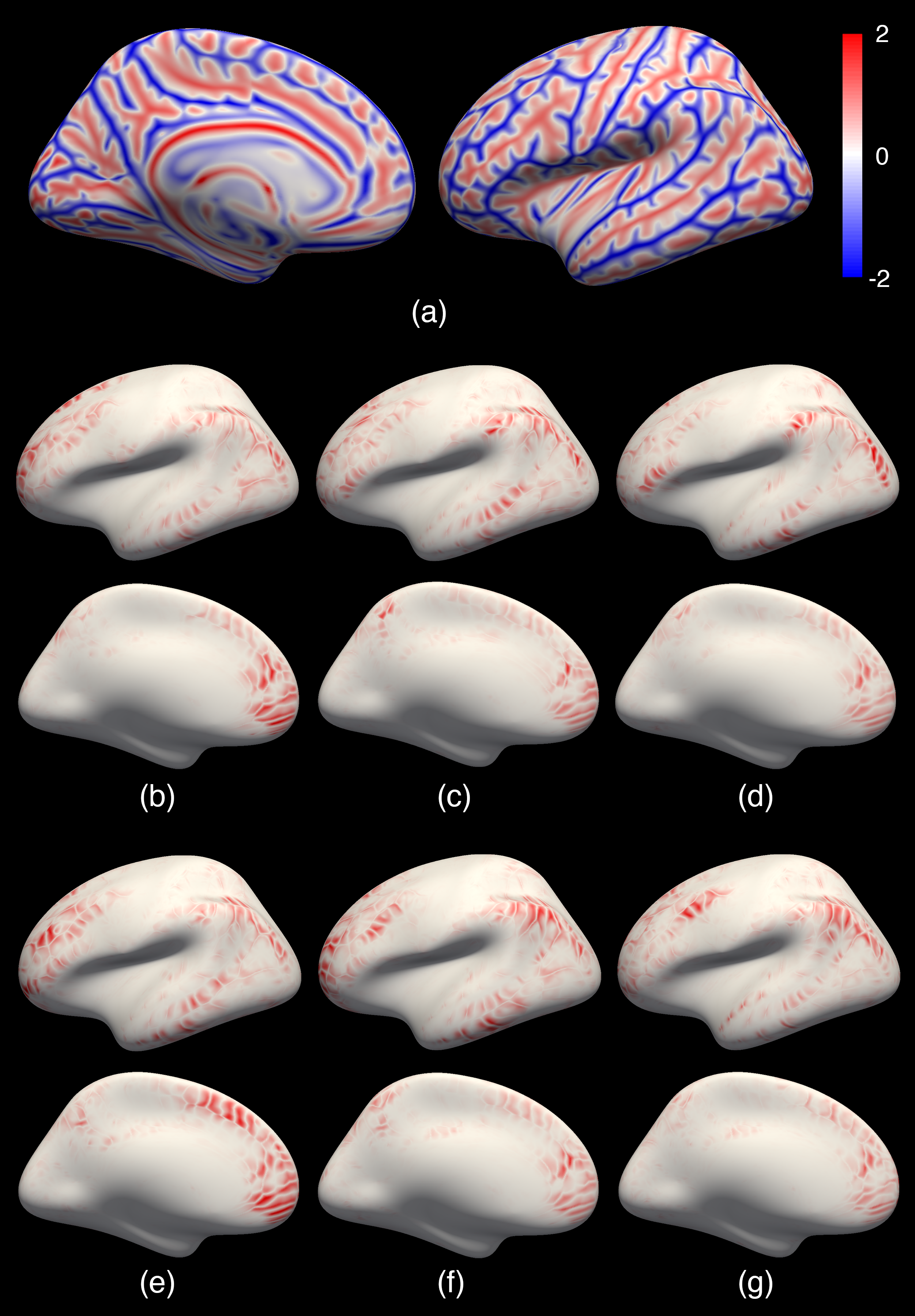}
  \caption{Learned geometric atlases from $6$ repeated training with randomly selected subjects. \textmd{(a) The mean atlas averaged over the $6$ runs. (b) - (g) The absolute difference between each run and the mean. Red indicates sulci and blue indicates gyri. The learned atlases are highly consistent across runs with all major folds nearly identical to each other and subtle differences in prefrontal, inferior parietal, and middle temporal areas.}}
  \label{fig:stability}
\end{figure}

\subsection{Atlas stability}
\label{sec:res:stability}
Fig.~\ref{fig:stability} shows the geometric atlases learned from the $6$ repeated training runs, learned identically in all respects except that the subjects and their orders are randomly selected as the input to the training. Fig.~\ref{fig:stability}~(a) shows the mean averaged over the $6$ runs and (b) - (g) shows the absolute difference between each run and the mean. These atlases are highly consistent across runs with all major folds nearly identical to each other and subtle differences in the prefrontal, inferior parietal, and middle temporal regions as reflected in the difference images. This strong consistency is quantitatively confirmed by the Pearson correlations $0.94\pm 0.01$ among all $15$ possible pairs of these atlases. The results support that the atlas construction and its learning procedure in our proposed JOSA method are stable and not biased towards any specific subject.

\begin{figure}[tb]
  \centering
  \includegraphics[width=0.8\linewidth]{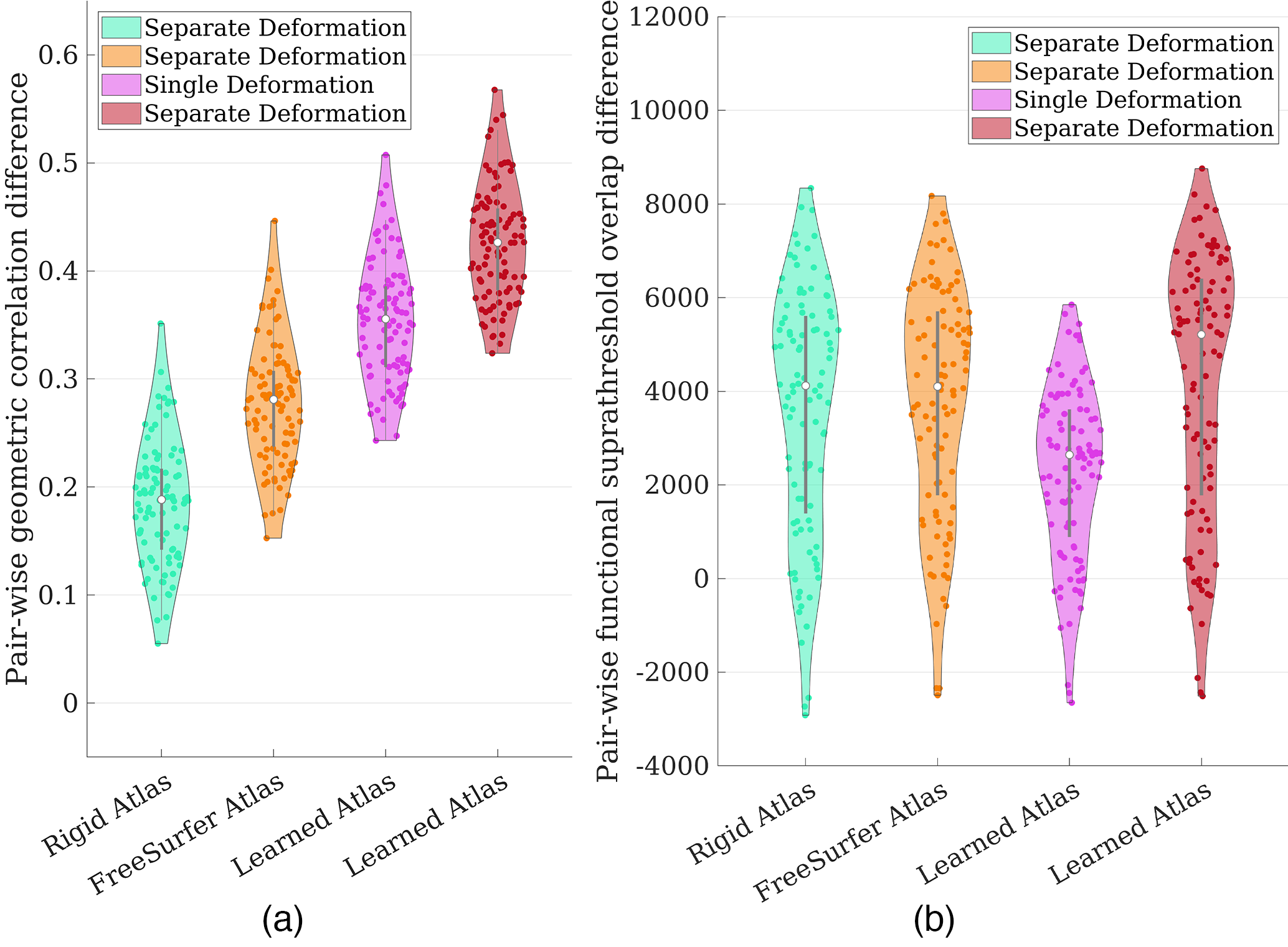}
  \caption{Violin plot of pair-wise difference between each registration method to rigid alignment for ablation studies. \textmd{(a) Geometric difference measured by correlation; (b) Functional difference measured by the suprathreshold overlaps. Geometric improvement is mainly attributable to a better atlas whereas functional improvement is primarily due to the separate modeling of the deformation fields.}}
  \label{fig:ablation}
\end{figure}

\subsection{Ablation study}
\label{sec:res:ablation}
Fig.~\ref{fig:ablation} shows the pair-wise differences from each registration method to the rigid result in the ablation study. The first and the second column shows the result using separate deformations for geometry and function as with JOSA but varying the atlas used as the registration target. The third column shows the result using a single deformation for both geometry and function but the same learned atlas as with JOSA. The last column shows the full JOSA result for easy reference. The substantial improvement in geometry is mainly attributable to a better atlas, whereas the improvement in the registration of function is primarily due to the separate modeling of deformation between geometry and function. We highlight the importance of this separate modeling of deformations. The third column suggests that the registration performance is sub-optimal for both geometry and function if they share the same deformation, which is commonly assumed (anatomy predicts function).

\section{Conclusion and discussion}
\label{sec:discussion}

In this work, we presented JOSA, a novel registration framework that jointly models the relationship between folding patterns and cortical function. In the early 1900s, the correspondence between the two was discovered in primary motor and sensory regions where a cortical homunculus was drawn to illustrate the distorted mapping of human body parts to sensorimotor functions \citep{Penfield_1937_B_SomaticMotor}. Similarly, the functional properties of primary visual cortex have been well studied, where the central and peripheral visual fields are located in the posterior and anterior part of the calcarine sulcus, respectively \citep{Dougherty_2003_JoV_VisualField,Sereno_1995_S_BordersMultiple,Tootell_1998_PNASU_FunctionalAnalysis,Engel_1997_CC_RetinotopicOrganization,Wandell_2007_N_VisualField,Engel_1994_N_FMRIHuman}, and the topographic mapping of the visual field to the visual cortex follows the complex-logarithm rule \citep{Schwartz_1980_VR_ComputationalAnatomy,Polimeni_2010_N_LaminarAnalysis}. However, the anatomo-functional mapping in higher-order functional areas, e.g., language, emotion, etc., is highly variable among individuals compared with the primary regions \citep{Steinmetz_1991_N_FunctionalAnatomy,Fischl_2008_CC_CorticalFolding,Frost_2012_N_MeasuringStructural,Robinson_2014_N_MSMNew}. This variability is mainly attributable to the non-unique optimal solutions for geometric registration, where, without the guidance from functional data, their performance can be sub-optimal in registration of function. For example, there are several secondary folds around the active language region near the frontal eye field. Aligning any of them may yield equally good geometric registration, but only one of them may be optimal for functional registration. This is not obvious without access and visualization of the functional data of each individual subject.

By using a semi-supervised training strategy with functional data as the guidance, JOSA is able to choose one appropriate deformation from many valid geometric solutions that produces optimal functional registration. Fig.~\ref{fig:violin} suggests that both JOSA and SphereMorph (no functional data) achieved comparable performance in geometric alignment, but only JOSA substantially improved functional registration. Moreover, a structure-function mismatch may exist and be different across individuals, i.e., different subjects may use slightly different regions to process similar function. By allowing small deviations of the functional alignment from geometric alignment, JOSA yields superior performance in registration of both folding patterns and functional properties relative to other traditional or learning-based methods. Fig.~\ref{fig:ablation}, the 3rd column showed that both the geometric and functional registration are sub-optimal when a single (shared between geometry and function) deformation is used.

\begin{figure}[tb]
  \centering
  \includegraphics[width=0.7\linewidth]{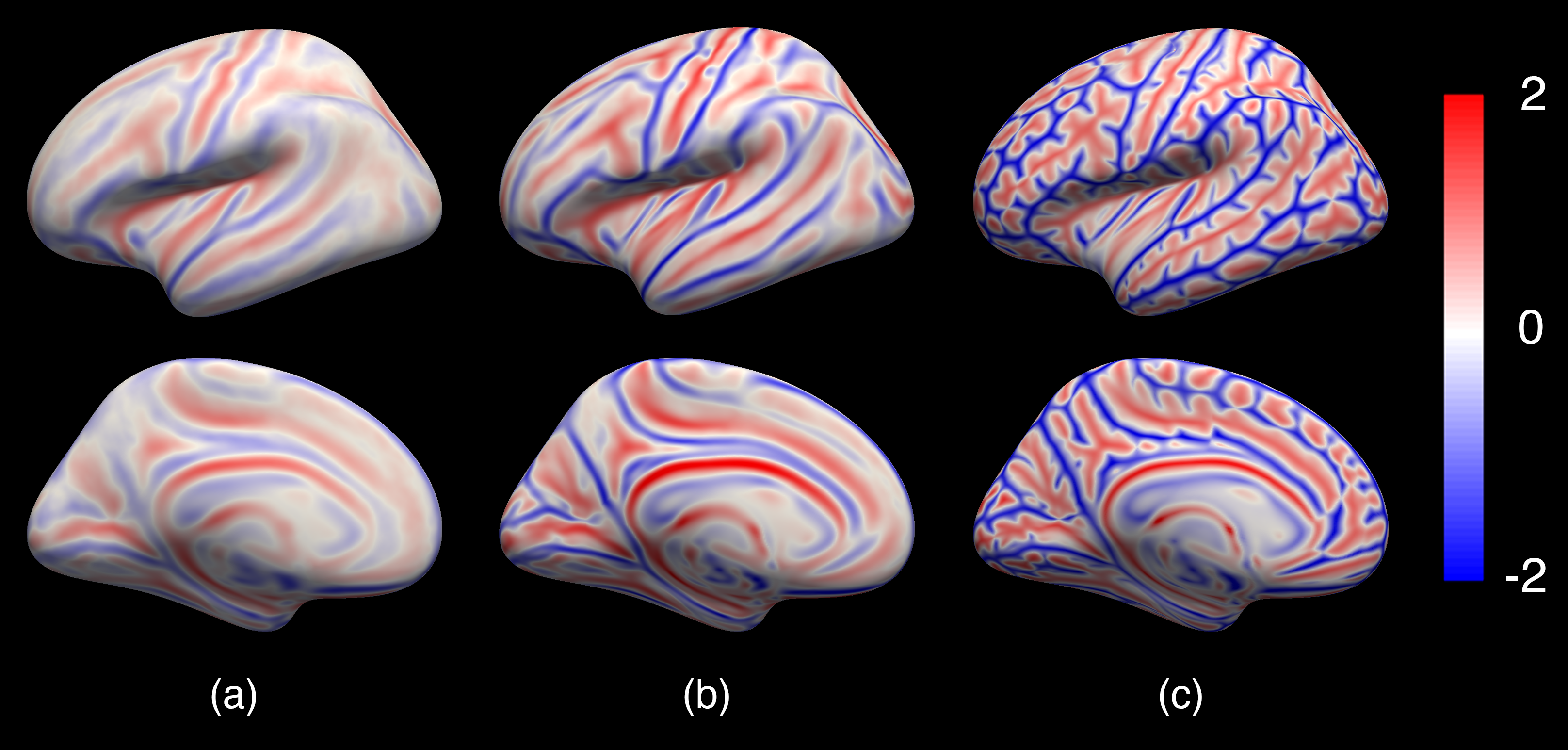}
  \caption{Comparison of atlases. \textmd{(a) Rigid atlas; (b) FreeSurfer atlas; (c) JOSA atlas. The learned atlas from JOSA provides more anatomical definition than rigid or FreeSurfer atlas, thus supporting registration with higher resolution and finer details.}}
  \label{fig:atlases}
\end{figure}

In contrast to many other learning-based methods, JOSA also estimates a population-specific atlas as the registration target during training. Fig.~\ref{fig:atlases} shows the learned atlas in (c) compared to atlases constructed by averaging over rigidly registered training subjects (a) and averaged FreeSurfer registered training subjects in (b). We find that the JOSA atlas provides more anatomical definition, supporting registration with higher resolution and finer details, which also contributes to the improved performance of JOSA (particularly in geometric registration).

Atlas or template construction has been widely studied in classical iterative approaches \citep{Collins_1995_HBM_Automatic3D,Fischl_1999_HBM_HighresolutionIntersubject,Desikan_2006_N_AutomatedLabeling,Destrieux_2010_N_AutomaticParcellation,Joshi_2022_JoNM_HybridHighresolution}. These methods build atlases by repeatedly registering subjects to an estimated atlas, and estimating a new atlas by averaging the registered subjects~\citep{Dickie_2017_FN_WholeBrain}. Inherited from this iterative construction approach, a bias towards the initial subject may be introduced to the atlas. JOSA, as one of the very recent learning-based methods, not only facilitates faster atlas construction \citep{Lee_2022_CiBaM_MulticontrastComputed,Lee_2022_MI2IP_SupervisedDeep}, but also potentially avoids the bias to any individual subject by random selection of subjects as inputs during training \citep{Cheng_2020_MUPPPIA_UnbiasedAtlas}, as illustrated in Fig.~\ref{fig:stability}. In this work, JOSA learned an atlas specific to the population that was used for training \citep{Lipkin_2022_SD_ProbabilisticAtlas}. It is, however, straightforward to extend this framework to learn any population-specific atlases, optionally conditioned on clinical attributes if desired \citep{Dalca_2019_ANIPS_LearningConditional,Dey_2021_PIICCV_GenerativeAdversarial,Ding_2022_PICCVPR_AladdinJoint}.

Using a semi-supervised training strategy, the deformation fields can be solely determined from geometric features, which can be derived from a single T1-weighted MR scan. Therefore, JOSA not only lifts the burden of acquiring functional data during inference, but also provides capacity for extension to other comprehensive data modalities that are related to geometry but difficult or even impossible to acquire, such as architectonics, transcriptomics, or proteomics. This geometry-only-based registration model further avoids the circularity issue in any subsequent analyses of functional or any other auxiliary data. In another words, if the functional data is used as one of the inputs to drive the registration, then any evaluation or analysis of the functional data after registration will be potentially biased as the network has observed the data before.

The current approach for better registration of brain function is limited by the use of one task fMRI localizer, the language localizer, used in this study. The partial coverage of the brain by the language-active regions may not provide useful registration guidance for functions that use different parts of the cerebral cortex. We therefore plan to extend JOSA to training with multi-modal datasets, enabling accurate registration of an array of brain functional responses. Although the language dataset used in this study \citep{Lipkin_2022_SD_ProbabilisticAtlas} is one of the largest task fMRI datasets publicly available, another caveat in the training of the registration network is the limited number of samples (subjects). Because the functional data were used only in loss evaluation through the semi-supervised block (i.e., they were not seen by the convolutional layers directly), it poses a significant challenge for the network to learn the anatomo-functional relationships, hence often resulting in an overfitting issue when the sample size is small. We also acknowledge the difficulty in data acquisition. One potential solution in the future is to use a pooled dataset from different task contrasts or even resting-state fMRI data, where subjects may have different functional data available, and the network can be trained dynamically using the corresponding available features, atlases, and loss functions.

Nevertheless, the superior performance of JOSA is mainly attributable to the modeling of the deviation between geometric and functional deformations. We hope this work elucidates the potential mismatch between brain geometry and function that might have been overlooked in the past and provides some new insights into the development of registration methods using joint analysis of brain structure and function in the future.

\section*{Acknowledgments}
Support for this research was provided in part by the BRAIN Initiative Cell Census Network grant U01MH117023, the National Institute for Biomedical Imaging and Bioengineering (R01EB023281, R01EB006758, R21EB018907, R01EB019956, P41EB030006), the National Institute on Aging (R56AG064027, R01AG064027, R01AG008122, R01AG016495, R01AG070988), the National Institute of Mental Health (UM1MH130981, R01MH123195, R01MH121885, RF1MH123195), the National Institute for Neurological Disorders and Stroke (R01NS052585, R21NS072652, R01NS070963, R01NS083534, U01NS086625, U24NS100591, R01NS105820, R21NS109627, RF1NS115268, U01NS132181), the National Institute on Deafness and Other Communication Disorders (R01DC016607, R01DC016950), the NIH Director's Office (DP2HD101400), the James S. McDonnell Foundation. This work was also made possible by the resources provided by Shared Instrumentation Grants S10RR023401, S10RR019307, and S10RR023043, the NIH Blueprint for Neuroscience Research (U01MH093765), part of the multi-institutional Human Connectome Project. Additional support was provided by the Amazon Fellowship from the Science Hub, administered by the MIT Schwarzman College of Computing, and the International Doctoral Fellowship from the American Association of University Women (AAUW). Much of the computation resources required for this research was performed on computational hardware generously provided by the Massachusetts Life Sciences Center (https://www.masslifesciences.com).

\bibliographystyle{model2-names.bst}
\bibliography{refs}

\section*{Supplementary material}

\renewcommand{\thefigure}{S\arabic{figure}}
\setcounter{figure}{0}

\begin{figure}[ht]
  \centering
  \includegraphics[width=0.8\linewidth]{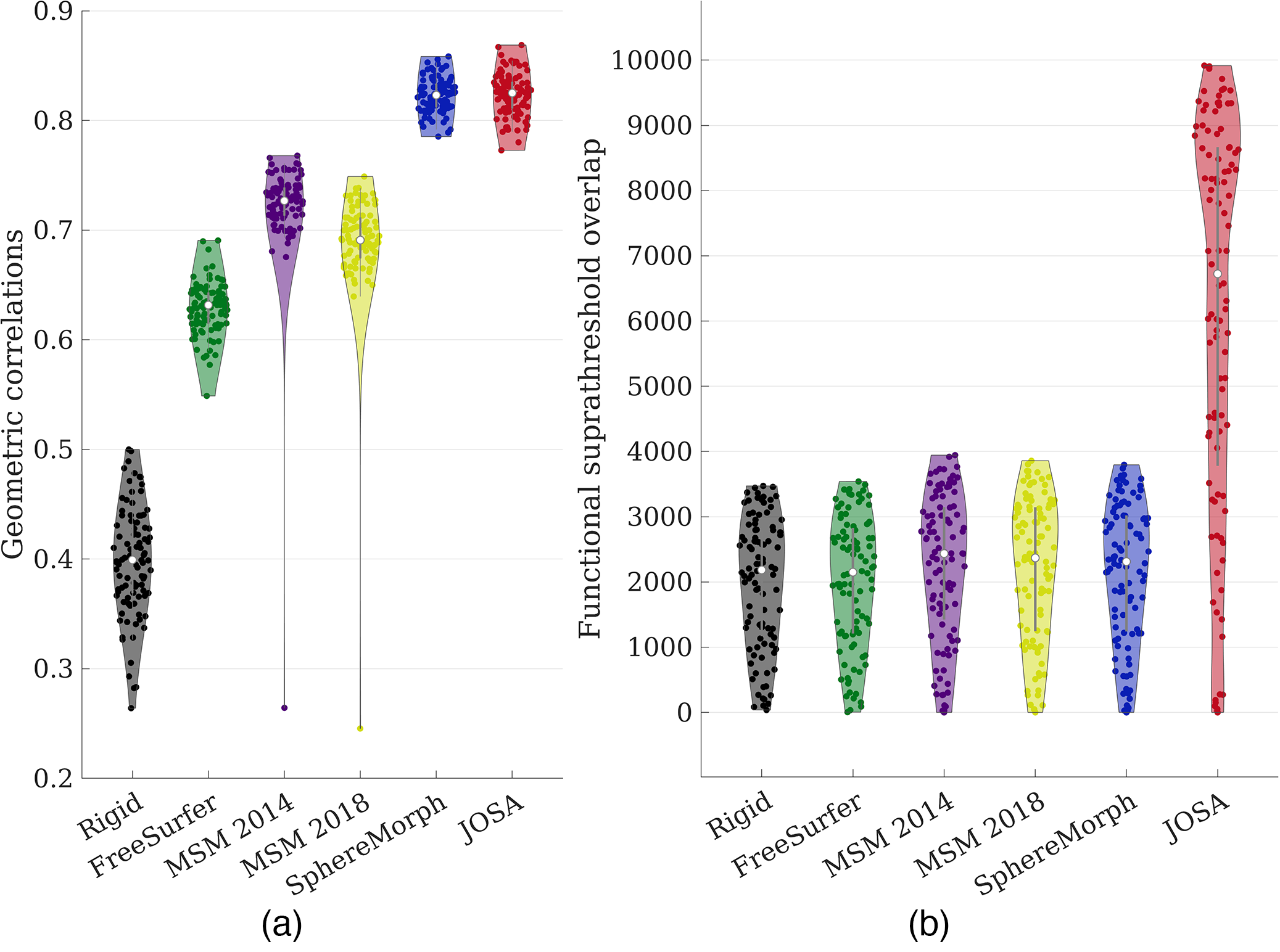}
  \caption{Violin plot of the original quantitative measures between individuals and group mean. \textmd{(a) Geometric correlations; (b) Functional overlaps.}}
  \label{fig:orig_violin}
\end{figure}

\begin{figure}[ht]
  \centering
  \includegraphics[width=0.55\linewidth]{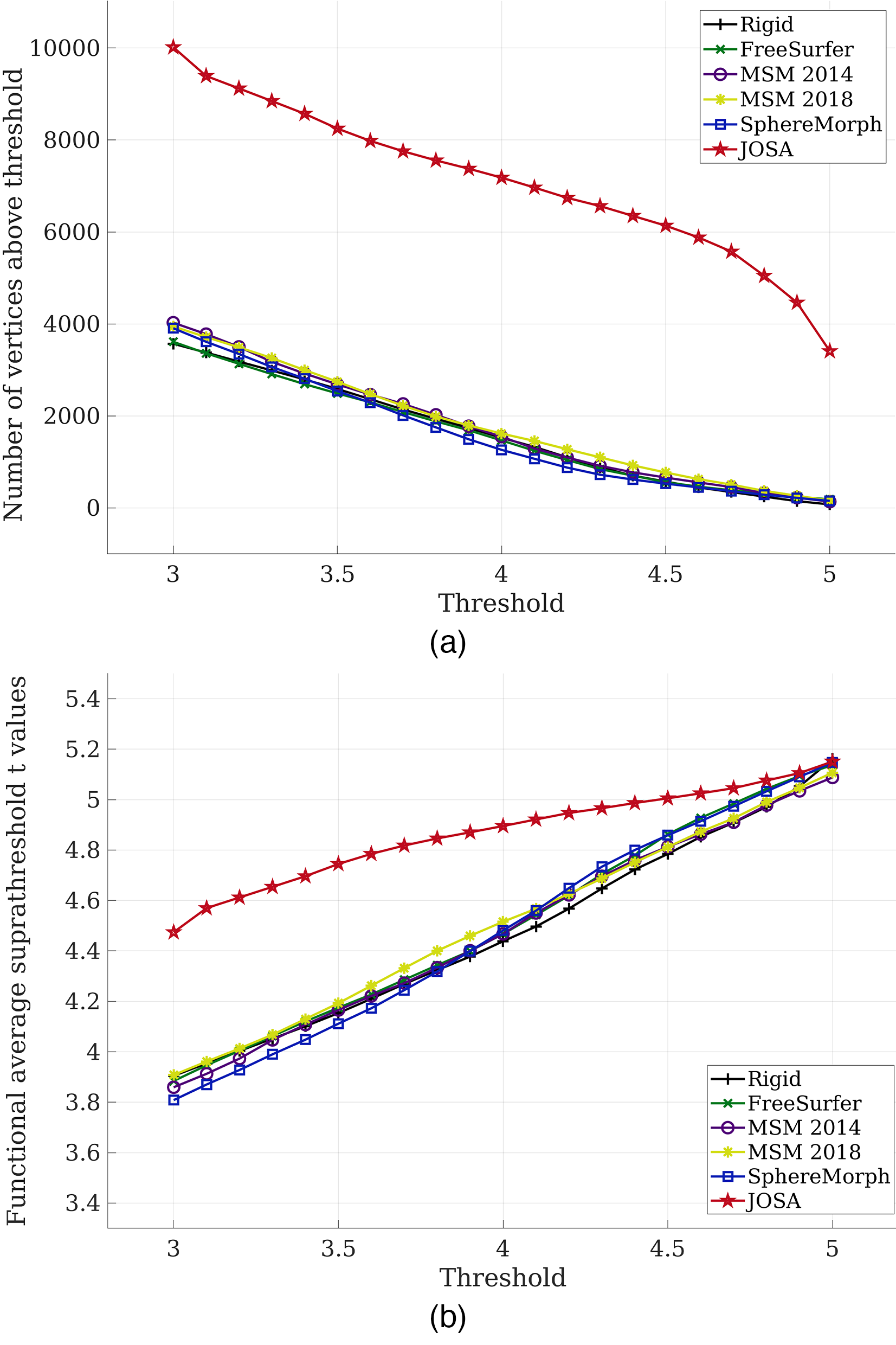}
  \caption{Impact factors of functional total suprathreshold t-values as a function of the threshold. \textmd{(a) Number of vertices within regions that have functional values above the threshold; (b) The averaged functional values within regions in (a).}}
  \label{fig:vert_strength}
\end{figure}

\end{document}